

\input harvmac
\input epsf
\noblackbox
\pageno=0\nopagenumbers\tolerance=10000\hfuzz=5pt
\line{\hfill CERN-TH/95-132}
\vskip 36pt
\centerline{\bf NEXT-TO-LEADING ORDER EVOLUTION OF}
\centerline{\bf STRUCTURE FUNCTIONS AT SMALL $x$ AND LARGE $Q^2$
                       \footnote{$^\star$}{Presented by RDB.}}
\vskip 36pt\centerline{Richard~D.~Ball\footnote{$^\dagger$}{On leave
{}~from a Royal Society University Research Fellowship.}
 and Stefano~Forte\footnote{$^\ddagger$}{On leave
{}~from INFN, Sezione di Torino, Italy.}}
\vskip 12pt
\centerline{\it Theory Division, CERN,}
\centerline{\it CH-1211 Gen\`eve 23, Switzerland.}
\vskip 60pt
{\medskip\narrower
\lineskiplimit=1pt \lineskip=2pt
\centerline{\bf Abstract}
\noindent
We show that a unified approach to the perturbative evolution of
structure functions which sums all logarithms of $Q^2$ and $1/x$
at leading and next-to-leading order yields results in full agreement
with the 1993 HERA data for $F_2$. This makes it possible to
determine $\alpha_s$ surprisingly accurately from these data alone.
}
\vfill
\centerline{Talk given at the}
\centerline{XXXth Rencontres de Moriond, {\it ``QCD and
High Energy Interactions''},}
\centerline{ Les Arcs, March 1995,}
\centerline{ to be published in the proceedings (Editions Fronti\`eres).}
\vskip 20pt
\line{CERN-TH/95-132\hfill}
\line{May 1995\hfill}
\vfill\eject
\footline={\hss\tenrm\folio\hss}
\nopagenumbers


\def\frac#1#2{{{#1}\over {#2}}}
\def\half{\hbox{${1\over 2}$}}

\def\smallfrac#1#2{\hbox{${{#1}\over {#2}}$}}

\def\GeV{{\rm GeV}}

\catcode`@=11 
\def\slash#1{\mathord{\mathpalette\c@ncel#1}}
 \def\c@ncel#1#2{\ooalign{$\hfil#1\mkern1mu/\hfil$\crcr$#1#2$}}
\def\lsim{\mathrel{\mathpalette\@versim<}}
\def\gsim{\mathrel{\mathpalette\@versim>}}
 \def\@versim#1#2{\lower0.2ex\vbox{\baselineskip\z@skip\lineskip\z@skip
       \lineskiplimit\z@\ialign{$\m@th#1\hfil##$\crcr#2\crcr\sim\crcr}}}
\catcode`@=12 

\def\DZP{\hbox{D0$'$}}

\def\PR{{\it Phys.~Rev.~}}

\def\NP{{\it Nucl.~Phys.~}}
\def\NPBPS{{\it Nucl.~Phys.~B (Proc.~Suppl.)~}}
\def\PL{{\it Phys.~Lett.~}}

\def\ZP{{\it Zeit.~Phys.~}}

\def\vol#1{{\bf #1}}\def\vyp#1#2#3{\vol{#1} (#2) #3}


\nref\ZEUS{ZEUS Collaboration, \ZP\vyp{C65}{1995}{379}.}
\nref\Hone{H1 Collaboration, \NP\vyp{B439}{1995}{471}.}
\nref\Summing{R.D.~Ball and S.~Forte, CERN-TH/95-1, {\tt hep-ph/9501231}.}
\nref\ad{T.~Jaroszewicz, \PL\vyp{B116}{1982}{291}\semi
         S.~Catani \& F.~Hautmann, \PL\vyp{B315}{1993}{157},
                  \NP\vyp{B427}{1994}{475}.}
\nref\EHW{R.K.~Ellis, F.~Hautmann and B.R.~Webber,
Cavendish-HEP-94/18, {\tt hep-ph/9501307}.}
\nref\DGPTWZ{A.~De~Rujula et al, \PR\vyp{D10}{1974}{1649}.}
\nref\DAS{R.~D.~Ball and S.~Forte, \PL\vyp{B335}{1994}{77}.}
\nref\Test{R.~D.~Ball and S.~Forte, \PL\vyp{B336}{1994}{77}.}
\nref\Mont{R.D.~Ball and S.~Forte, \NPBPS\vyp{39B,C}{1995}{25}.}
\nref\lxalf{A.D.~Martin et al, \PL\vyp{B266}{1991}{173}\semi
M.~Virchaux and A.~Milsztajn,\PL\vyp{B274}{1992}{221}.}

At small $x$ the structure function $F_2^p(x,Q^2)$ measured recently at
HERA \ZEUS\Hone, depends in effect
on two scales, $Q^2$ and $s=\big(\smallfrac{1-x}{x}\big)Q^2$, instead
of just the single scale $Q^2$ relevant at large $x$. The
renormalization group equations which govern the
perturbative evolution of singlet parton distribution functions
thus describe evolution in both $s$ and $Q^2$ (or
equivalently $1/x\simeq s/Q^2$ and $t\equiv\ln Q^2/\Lambda^2$), and
should sum up all leading (and subleading) logarithms of both $Q^2$
and $s$ (or $Q^2$ and $1/x$). In fact it is possible to construct
unified perturbative evolution equations for leading twist singlet
parton distribution functions which are straightforward
generalizations of the usual singlet Altarelli-Parisi equations,
but with new
splitting functions which explicitly incorporate all the logarithms
of $1/x$. The resulting equations\Summing\ may be used to calculate
perturbative  evolution down to
arbitrarily small values of $x$, provided only that $Q^2$ is
sufficiently large that higher twist effects may be ignored. All
nonperturbative effects (at both large and small $x$) are then
factorized into the initial condition at $t_0$.

The singlet splitting functions in the the double leading expansion
scheme, which treats the two scales symmetrically (see ref.\Summing),
take the form $P^{ij}(x,t) = P^{ij}_{\rm LO}(x)+
\smallfrac{\alpha_s(t)}{2\pi}P^{ij}_{\rm NLO}(x,t)+\cdots$, where
$P^{ij}_{\rm LO}(x)=P^{ij}_1(x)+P^{ij}_s(x,t)$,
$P^{ij}_{\rm NLO}=P^{ij}_2(x)+P^{ij}_{ss}(x,t)$, $P^{ij}_1(x)$ and
$P^{ij}_2(x)$ are the usual one and two loop splitting functions,
while $P^{ij}_s(x,t)$ and $P^{ij}_{ss}(x,t)$ are
(convergent) sums of leading and subleading singularities
respectively: for $x>x_0$ they are all zero, while for $x<x_0$
$P^{qg}_s(x,t)=P^{qq}_s(x,t)=0$, but
\eqn\splitsing{\eqalign{
xP^{gg}_s(x,t)&=\smallfrac{C_A}{C_F}xP^{gq}_s(x,t)
=2C_A\sum_{n=4}^{\infty}a_n\smallfrac{1}{(n-1)!}
\lambda_s(t)^{n-1}\xi^{n-1},\cr
xP^{qg}_{ss}(x,t)&=\smallfrac{C_A}{C_F}xP^{qq}_{ss}(x,t)
=\smallfrac{2}{3}T_R n_f\sum_{n=3}^{\infty}
\tilde a_n\smallfrac{1}{(n-2)!}\lambda_s(t)^{n-1}\xi^{n-2},\cr
xP^{gg}_{ss}(x,t)&=\smallfrac{C_A}{C_F}xP^{gq}_{ss}(x,t)=
2C_A\sum_{n=3}^{\infty} b_n
\smallfrac{1}{(n-2)!}
\lambda_s(t)^{n-1}\xi^{n-2},\cr
}}
where $\lambda_s(t)\equiv 4\ln 2\smallfrac{C_A}{\pi}\alpha_s(t)$,
$\alpha_s(t)$ is the two-loop running coupling, and
$\xi=\log\smallfrac{x_0}{x}$.

The coefficients $(a_n,\tilde a_n)$ may
be found in ref.\Summing: they are computed using expressions derived in
ref.\ad. The coefficients $b_n$ (and the color-charge relation between
$P^{gg}_{ss}$ and $P^{gq}_{ss}$) can be fixed uniquely at NLO
by requiring momentum conservation: $b_n = -a_n - \smallfrac{T_R n_f}{3
C_A}\tilde a_n$. This fixes the
arbitrariness implicit in alternative schemes in which momentum is
conserved separately for quarks and gluons\EHW, and furthermore
makes the explicit computation of $P^{gg}_{ss}$ and
$P^{gq}_{ss}$ unnecessary at NLO: subleading corrections to the BFKL
kernel would only be necessary for NNLO computations.
The parameter $x_0$, which marks the boundary between large and small
$x$ regions, must eventually be fixed empirically: we vary it in the
range $0\leq x_0\leq 0.1$, considering larger values of $x_0$ to be
unreasonable.

Each of the series for the splitting functions is
uniformly convergent on any closed
interval of $x$ which excludes the origin, and can thus be used to
extend perturbative evolution down to arbitrarily small values of
$x$\Summing. It follows that their Mellin transforms (the anomalous
dimensions $\gamma_N$) have no physically relevant singularities
beyond those present at fixed order in $\alpha_s$ (so in particular
any cuts generated by all order resummations may in practice be
ignored).

The asymptotic behaviour of the parton distribution functions, as
solutions to these unified evolution equations, depends on the
particular limit adopted in the $x$-$t$ plane.
In the Bjorken limit $t\to\infty$, with $x$ fixed, the most important
parts of the splitting functions are $P_1$ to LO and $P_2$ at NLO.
The dependence on $x$ is essentially nonperturbative, being given by the
initial condition at $t_0$, while the dependence on $t$ is
perturbative. In the Regge
limit $\xi\to\infty$ at fixed $t$, the most important parts of
the splitting functions are the leading and subleading singularities
$P_s$ and $P_{ss}$. In this limit $F_2$
eventually rises as $x^{-\lambda(t)}$, the power $\lambda(t)$ being
given by Regge theory for $t\sim t_0$, but calculable perturbatively
for $t$ sufficiently large ($\lambda(t)=\lambda_s(t)$)\Summing.
Finally, in the double scaling limit $\xi\to\infty$,
$\zeta\equiv\ln\smallfrac{t}{t_0}\to\infty$, $\xi/\zeta$ rising as
some power of $\zeta$, but slower than an exponential, the asymptotic
form of singlet parton distribution functions may be
determined completely in perturbation
theory\DGPTWZ\DAS\Summing. In particular the structure function $F_2(x,t)$
then grows as $R_F\inv\equiv{\cal N}\sigma^{-\half}
\rho\inv e^{2\gamma\sigma-\delta\zeta}$, where $\sigma\equiv
\sqrt{\xi\zeta}$,
$\rho\equiv\sqrt{\xi/\zeta}$, $\gamma = \smallfrac{6}{5}$ and $\delta
= \smallfrac{61}{45}$. This leads to a double asymptotic
scaling in $\sigma$ and $\rho$ which is indeed seen experimentally
\DAS\Test (see fig.~1). In the double scaling limit (and thus in the kinematic
regime studied at HERA) the most important parts of the splitting
functions are thus the pivotal leading singularity of $P_1$ (which
gives $\gamma$) and the subleading term (which gives $\delta$) at LO,
and the leading singularity of $P_2$ at NLO.

Two loop calculations of $F_2^p$ in the double scaling region were
performed in ref.\Mont: here we give similar calculations in the
double leading scheme. The solution of the NLO evolution equations
is straightforward, although care must be
taken to linearize all subleading corrections, to avoid spurious
sub-subleading terms in the solution. As in \Mont\ we use as
input the MRS distribution \DZP\
with a small-$x$ tail parameterized by the single parameter
$\lambda(Q_0)$ at the starting scale $Q_0$. The two parameters
$\lambda(Q_0)$ and $\alpha_s$ are then fitted by minimising the
total $\chi^2$ of the evolved distribution to the HERA data. The
resulting best-fit parameters are given in table~1, and the
corresponding distributions displayed in figure~1, in the form of
two scaling plots\DAS. The two loop curve (i.e. that with $x_0=0$)
now fits the data significantly
better than the scaling curve (which has $\chi^2 = 71/121$),
essentially because the most singular part of the two loop correction
reduces the slope of the $\sigma$ plot a little at lower values of
$\sigma$, a tendency which is evident in the data. The curve with
$x_0=0.1$ is almost identical to that with $x_0=0$, differing only at
large values of $\rho$; as explained in ref.\Summing, most of the
effect of the higher order logarithmic singularities can be absorbed
by suitably adjusting the boundary condition, either by changing $Q_0$ at fixed
$\lambda$, or (as here) $\lambda$ at fixed $Q_0$. If one were to
keep the boundary condition fixed one might instead have
the impression that the higher loop singularities have a large effect
on $F_2$ in the HERA region.

\topinsert\null\vskip -3.4truecm\hbox{\hskip -2.1truecm
\epsfxsize=11.truecm\epsfbox{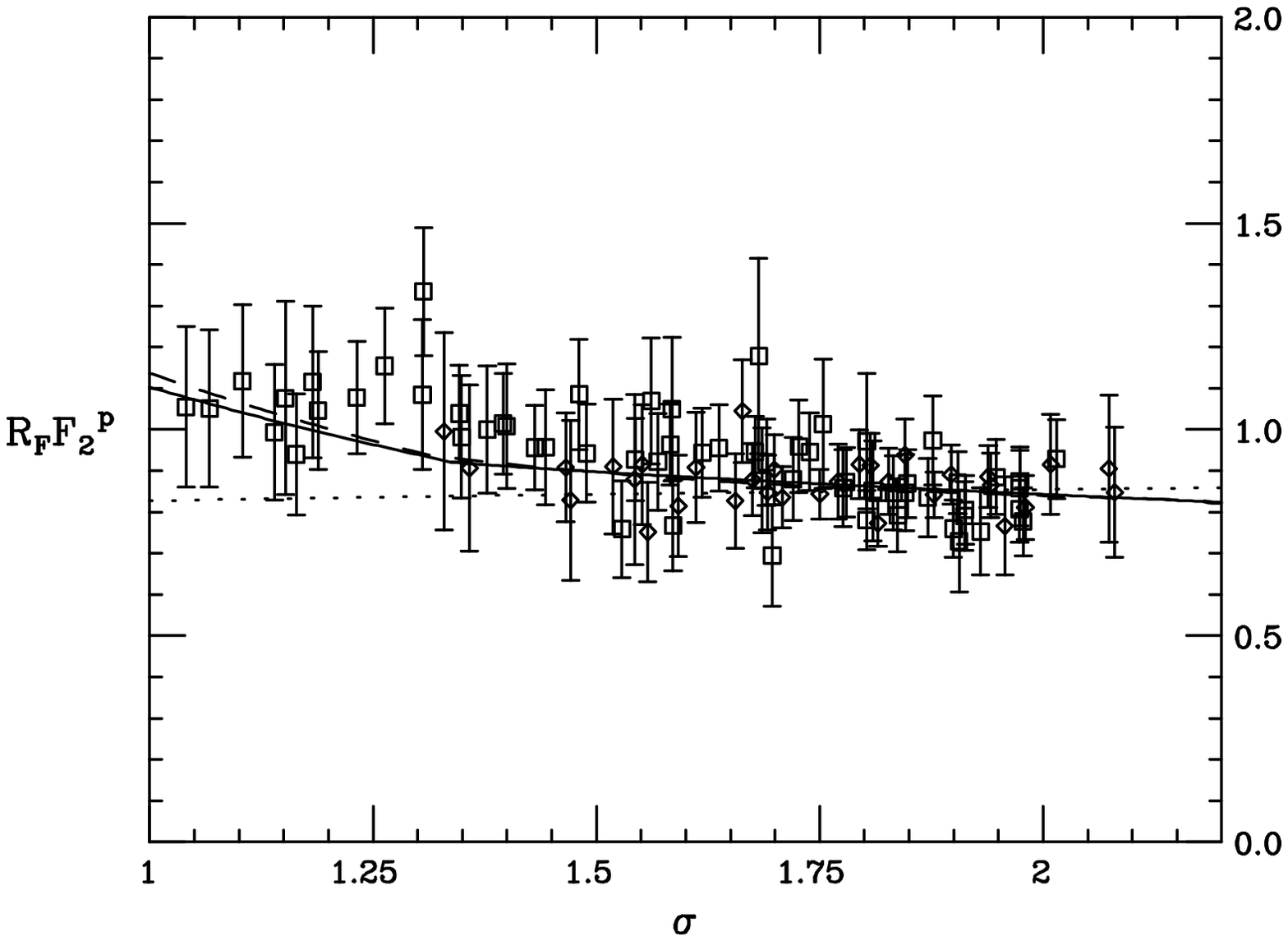}
\epsfxsize=11.truecm\epsfbox{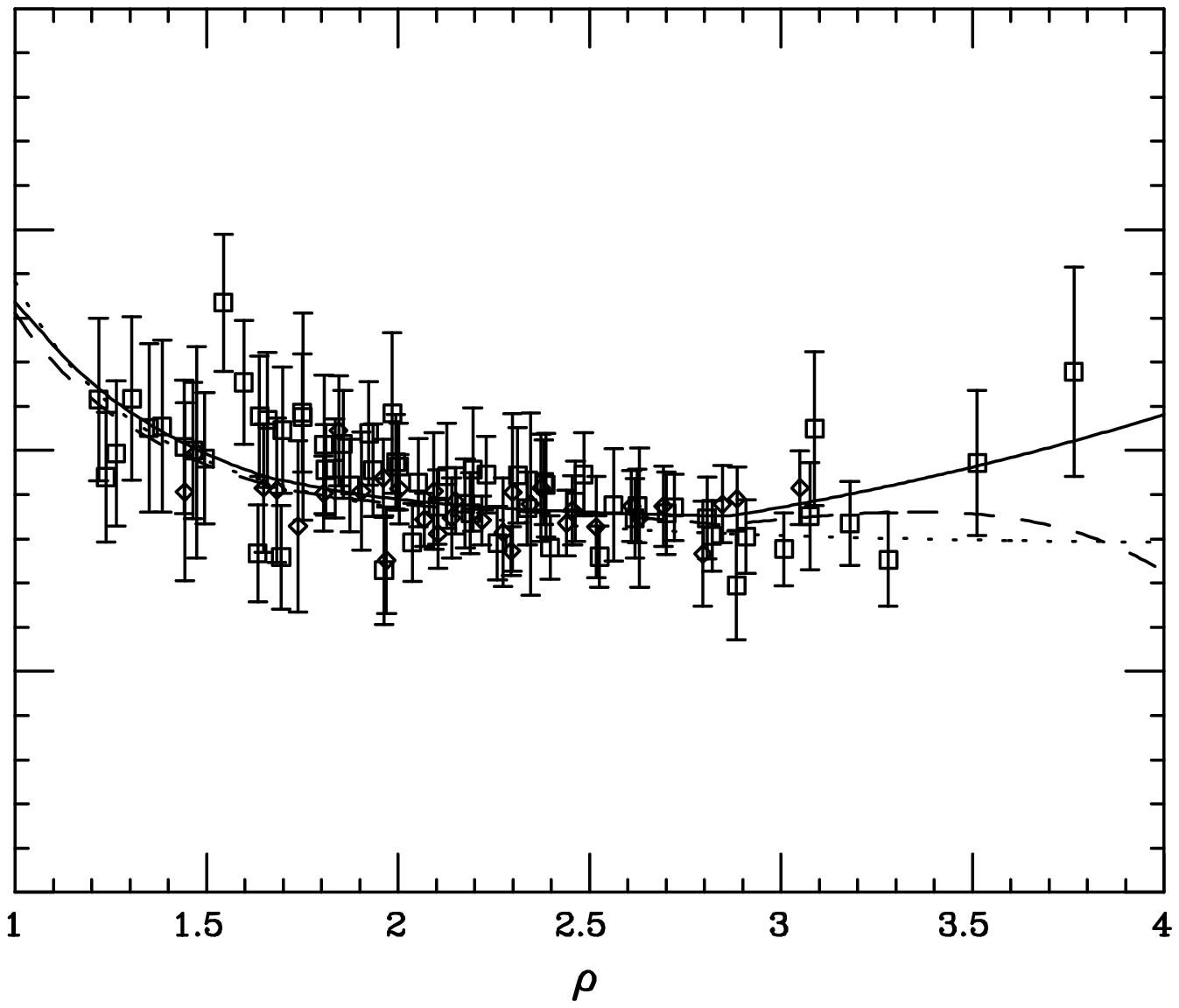}}
\vskip -3.5truecm
\centerline{\vbox{\hsize= 300pt \noindent\footnotefont
\baselineskip\footskip
Figure 1: Scaling plots of $R_F F_2^p$ vs $\sigma$ and $\rho$. All
1993 HERA data passing the cuts
$\sigma,\rho >1$, $Q^2 < (2 m_b)^2$ are included, and $R_F$ is
thus evaluated with $n_f=4$. The diamonds and squares are ZEUS\ZEUS\
and H1\Hone\ data, renormalized by $1.02$ and $0.95$ respectively.
The curves
are the simple double scaling \hbox{prediction \DAS} (dotted), and two
NLO double leading calculations with extremal values of
$x_0$: $x_0=0$ (i.e. two loops only) (solid) and $x_0=0.1$ (dashed).
The curves in the $\sigma$-plot have $\rho=2.2$: those on the
$\rho$-plot $\sigma=1.7$.
}}
\medskip\endinsert

\midinsert\hfil
\vbox{\footnotefont\baselineskip\footskip
      \tabskip=0pt \offinterlineskip
      \def\tablerule{\noalign{\hrule}}
      \halign to 350pt{\strut#&\vrule#\tabskip=1em plus1em
                   &\hfil#\hfil&\vrule#
                   &#\hfil&\vrule#
                   &#\hfil&\vrule#
                   &#\hfil&\vrule#
                   &\hfil#&\vrule#\tabskip=0pt\cr\tablerule
             &&\omit\hidewidth $x_0$\hidewidth
             &&\omit\hidewidth norms\hidewidth
             &&\omit\hidewidth $\lambda(2~\GeV )$\hidewidth
             &&\omit\hidewidth $\alpha_s(M_Z)$\hidewidth
             &&\omit\hidewidth $\chi^2$\hidewidth&\cr\tablerule
   &&  $0$  && $102\%\qquad 95\%$
              && $-0.23\pm 0.03$ && $0.122\pm 0.002$ && $58.4/120$ &\cr
   && $0.1$ && $99\%\qquad 93\%$
              && $-0.06\pm 0.06$ && $0.114\pm 0.004$ && $64.9/120$ &\cr
\tablerule
}}\hfil\medskip
\centerline{\vbox{\hsize= 300pt \noindent\footnotefont
\baselineskip\footskip
Table: Fitted parameters and $\chi^2$ for the two fitted curves in
figure~1. Statistical and systematic errors for each data have been
added in quadrature, and the normalizations of the two experiments
fitted within their stated uncertainties of $\pm 3.5\%$ and $\pm
4.5\%$ respectively\ZEUS\Hone.
}}
\medskip
\endinsert

Whereas the higher order singularities only affect $F_2$ close to the
nonperturbative boundary at $t=t_0$, they can have a relatively
large effect on the size of the quark and gluon distributions
at small $x$ and large $t$. To search for this effect (and, if it is found, to
measure $x_0$) it would be very interesting to have an independent
measurement of the gluon distribution, either through $F_L$ or heavy
quark production. If $x_0$ is as large as $0.1$, such a measurement
should find a significantly smaller gluon distribution than that
expected at two loops.

\topinsert\null\vskip-1.5truecm
\hbox{
\hskip .7truecm
\epsfxsize=6.7truecm\epsfbox{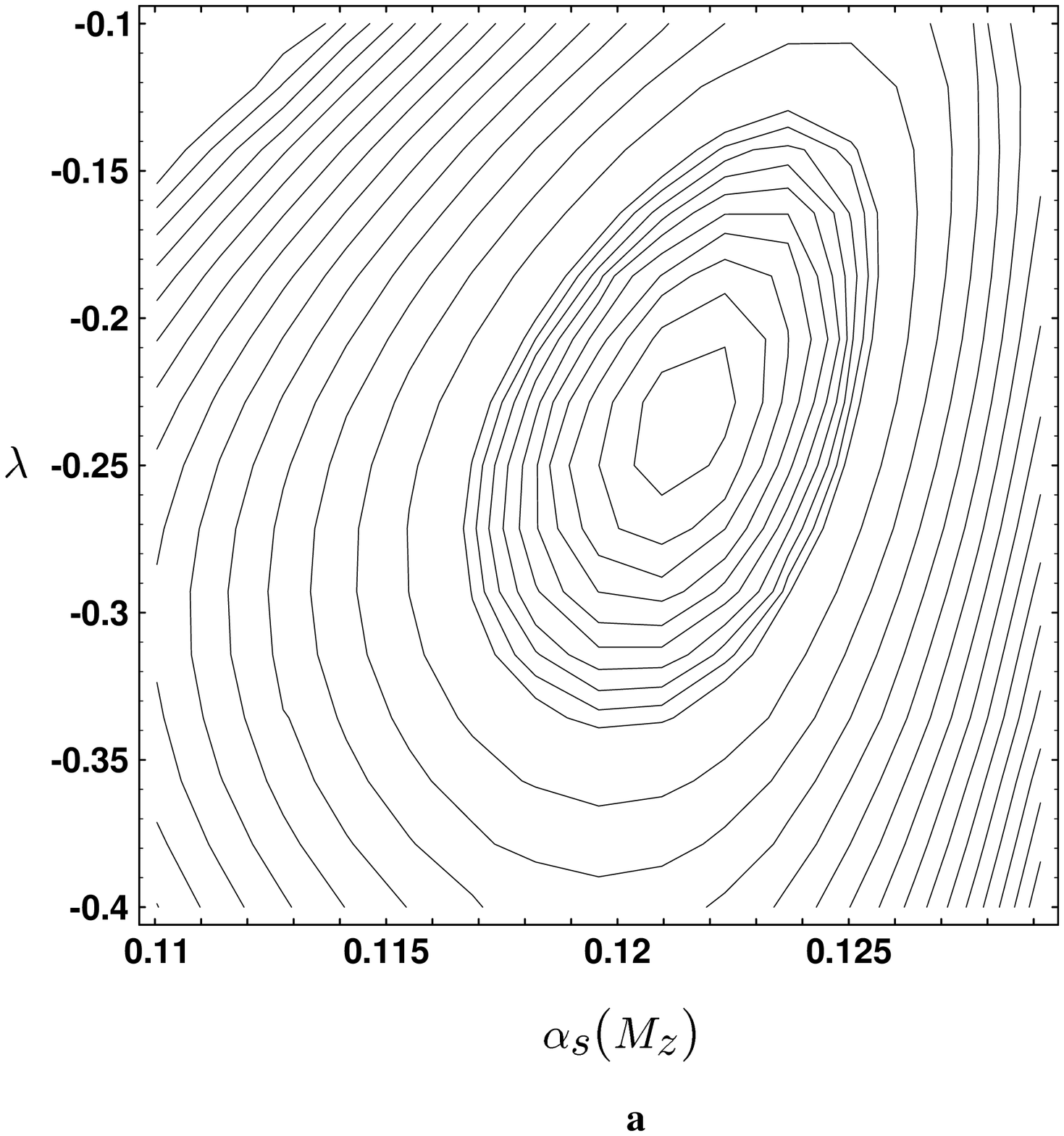}
\hskip 1.2 truecm
\epsfxsize=6.7truecm\epsfbox{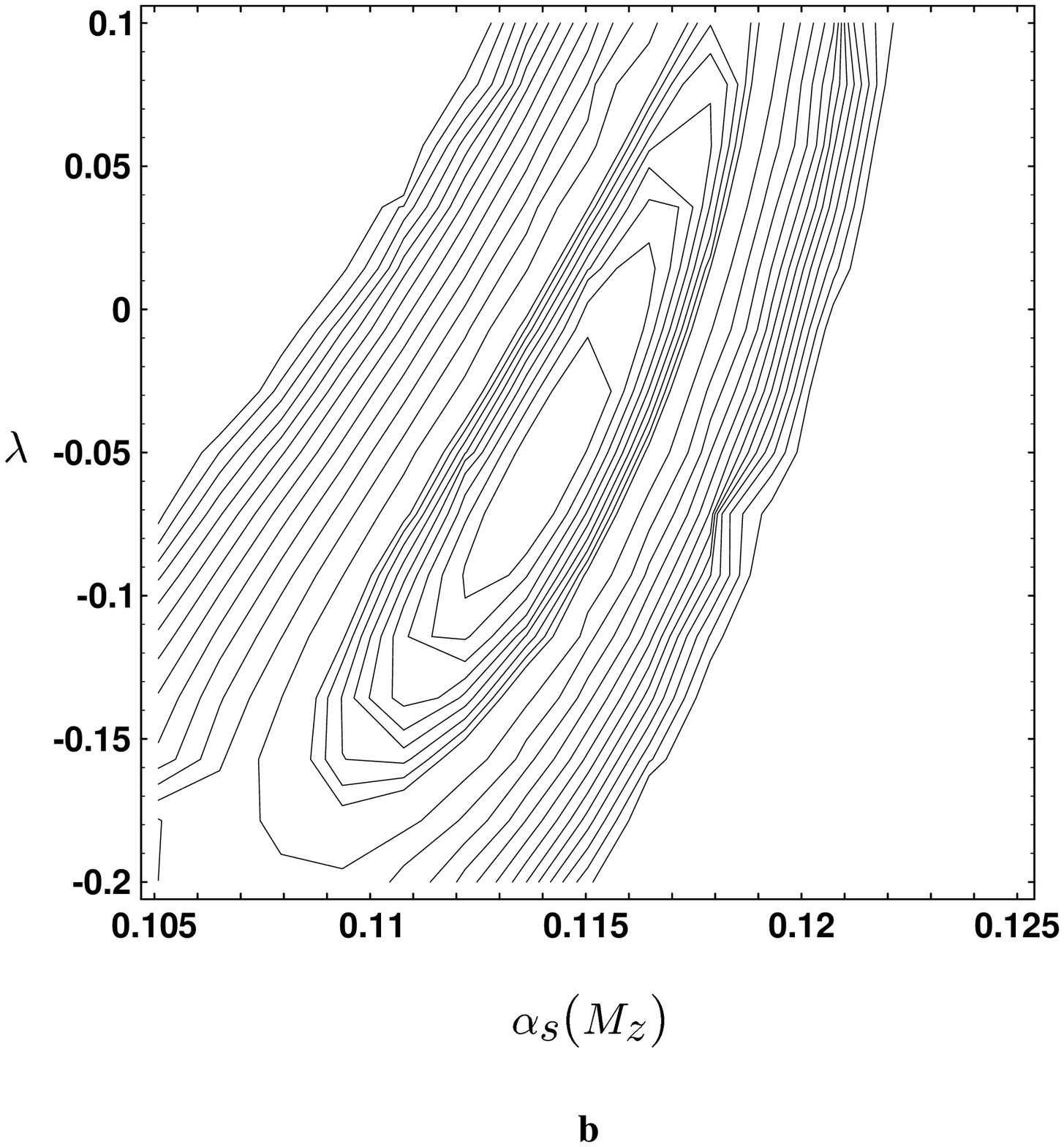}}
\centerline{\vbox{\hsize= 300pt \noindent\footnotefont
\baselineskip\footskip
Figure 2: Contour plots of the $\chi^2$ in the $\lambda$-$\alpha_s$
plane, for the two fits shown in the table: a) $x_0=0$; b) $x_0=0.1$.
Contours after the first
ten are at intervals of five units. }}
\medskip\endinsert

Since the rise in $F_2$ at small $x$ and large $Q^2$ is being driven
essentially by the triple gluon vertex, it depends rather strongly on
the value of $\alpha_s$. It should thus be a good place to measure
$\alpha_s$, requiring far lower statistics than at large $x$, where
one has to search instead for small violations of Bjorken scaling\lxalf.
Indeed the 1993 HERA data are already sufficient for such a
determination, as is apparent from the two parameter fits of shown in
the table (see also fig.~2), and the fact that the two loop
correction can be seen in the data. A preliminary error analysis gives
\eqn\alfa{\alpha_s(M_Z) = 0.120\pm 0.005{\rm (exp)}
\pm 0.010 {\rm (th)}:}
a more detailed determination will be presented elsewhere. As at large
$x$ the theoretical error is dominated by the renormalization scale
uncertainty, with an additional uncertainty here due to the unknown value
of $x_0$. Higher twist effects however seem to be very small.

\bigskip\noindent
{{\bf Acknowledgements:} We would like to thank
S.~Catani, M.~Ciafaloni and R.K.~Ellis for discussions.
}


\bigskip
\listrefs
\end